\documentstyle[epsf]{article}
\begin{document}

\title{Observational constraints for power-law spectra of 
density perturbations} 
\author{N.A. Arkhipova$^1$, V.N.Lukash$^2$, E.V.Mikheeva$^2$\\ 
$^1$Moscow State University, Vorobjovy Gory, 119899 Moscow\\ 
 $^2$Astro Space Center of Lebedev Physical Institute of RAS,\\ 
     Profsoyuznaya 84/32, 119810 Moscow}

\maketitle

\centerline{\bf Abstract}
We present an analysis of cosmological models with mixed dark 
matter (spectrum slope of density perturbations $n_S \in [0.6, 
1.6]$, hot dark matter abundance $\Omega_\nu \in [0.0, 0.6]$) in
spatially flat Universe, based 
on Press-Schechter formalism (P\&S). We found that the models with 
$n_S > 1$ and $\Omega_\nu < 0.5$ are preferable. Additionally, 
for all considered models we simulated $\Delta T/T\vert_{10^0}$ and 
compared it with the COBE measured amplitude of CMB anisotropy. We 
found that the models favourable for galaxy cluster abundance need 
large amount of cosmological gravitational waves (T/S $> 1$).

\section{Introduction}
A reconstruction of density perturbation spectrum is a key 
problem of the modern cosmology. It made a dramatic turn after 
detecting the primordial CMB anisotropy by DMR COBE (Smoot et 
al\cite{1}, Bennet et al\cite{2}) as the signal found at 
$10^0$ $\Delta T/T = 1.06 \times 10^{-5}$ appeared to be few 
times more than the expectable value of $\Delta T/T$ in the most 
simple and developed cosmological model -- standard CDM one 
(SCDM\footnote {$\Omega_M = 1$, $\Omega_b = 0.06$ (Walker at 
al\cite{3}), $\Omega_{CDM} = 0.94$, $h = 0.5$, no 
cosmological gravitational waves.}).

Currently there are a lot of experimental data (such as the
spatial distributions of galaxies, clusters of galaxies and  
quasars, bulk velosities, CMB anisotropy, and others) which can 
be used to reconstruct the density perturbation spectrum. 
Characteristic scales of data are different and vary from $\sim 
10$ Mpc which is a scale of nonlinearity to the horizon scale. 
However, it seems now the most crucial tests are large-scale CMB 
anisotropy and the number of galaxy clusters in {\it top-hat} 
sphere with radius $R = 8 h^{-1}$ Mpc = 16 Mpc. The former can 
be easy related to the amplitude of density perturbations 
through the SW effect (Sachs \& Wolfe\cite{4}):
$$
\frac{\Delta T}{T}(\vec e) = 
\frac{H_0^2}{2(2\pi)^{3/2}} 
\int \limits_{-\infty}^{\infty} \frac{1}{k^2} \delta_{\vec k} 
e^{i\vec k\vec x}d^3\vec k,\;\;\;\;
\vec x \simeq \frac{2\vec e}{H_0},
$$
where $\delta_{\vec k}$ is a Fourrier transform of density 
contrast $\delta(\vec x) \equiv \delta \rho / \rho$, $H_0$ is 
the Hubble constant, $\langle \delta_{\vec k} \delta_{\vec k'} 
\rangle = P(k) \delta(\vec k -\vec k')$, $P(k) = A k^{n_S} T^2 
(\Omega_\nu,k)$ is a power spectrum of density perturbations, $A
$ is the normalization constant, $T(\Omega_\nu,k)$ is a transfer
function. The latter determines the value of biasing parameter 
$b^{-1} \equiv \sigma_R$ for spatially flat Universe:
$$
\sigma_R^2 = \frac{1}{(2\pi)^3}\int_{-\infty}^\infty P(k) 
W^2(kR)d^3 \vec k,
$$
where $W(kR) = \frac{3}{(kR)^3}(\sin kR - kR\cos kR)$ is the 
Fourrier transform of hop-hat window function.

Obviously, both normalizations are model-dependent, the $\Delta 
T/T$ normalization depends on the amplitude of cosmological 
gravitational wave spectrum on large scale and, therefore, is 
related to the model of inflation, the $\sigma_R$ normalization 
does depend on the nature of dark matter. Here we prefer to fix
$\sigma_{16}$ to consider the relative
contribution of gravitational waves at COBE scale T/S as an
additional calculable parameter (instead of considering some
inflationary model).

Below, we report results based on P\&S formalism which deals with 
abundance of gravitationally bounded halos of dark matter.

\section{P\&S formalism}
P\&S formalism (Press \& Schechter\cite{5}) is built up on two 
assumptions: firstly, the density contrast $\delta$ is a 
Gaussian random field, and, secondly, the mass distribution of 
virialized halos of dark matter is the same as the distribution 
of high density peaks.
\begin{figure}[t]
\epsfxsize=9cm
\centerline{\epsfbox{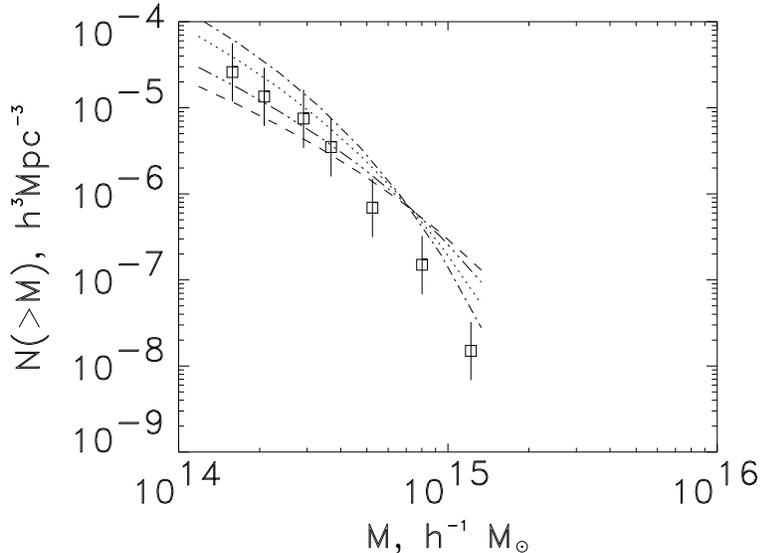}}
\caption{ 
Theoretical curves for 4 cosmological models: 
$n_S = 1.4$, $\Omega_\nu = 0.1$ (dot-dashed line); 
$n_S = 1.0$, $\Omega_\nu = 0.1$ (dotted line); 
$n_S = 1.4$, $\Omega_\nu = 0.5$ (3 dot-dashed line); 
$n_S = 1.0$, $\Omega_\nu = 0.5$ (dashed line). 
All models are normalized by $\sigma_{16} = 0.57$.
Empty squares with error bars are observational points taken
from Bachall \& Cen${}^{8)}$. }
\end{figure}

Since the density contrast smoothed by a window function $\delta
(\vec x,R)$ is the Gaussian random field too, we can easily 
calculate the fraction of space points at which the density 
exceeds any given value. Thus, the fraction of the points 
surrounded by a sphere of radius $R$ within which the density 
contrast exceeds $\delta_c$, is given by
$$
P(\delta > \delta_c) = \frac 1{\sqrt{2\pi} \sigma} \int 
\limits_{\delta_c}^\infty e^{-\frac{\delta^2}{2\sigma^2}}d 
\delta.
$$

According to P\&S this fraction can be identified with the
fraction of halos with mass exceeding $M_c = \frac{4\pi}3 \rho_0 
(1 + \delta_c) R^3$, where $\delta_{c}= 1.69$ (Gunn \& 
Gott\cite{6}). The integral mass distribution of nonlinear lumps 
is the following: 
$$
N( >M) 
= \int_M^{+\infty} \frac{dn}{dM} dM 
= \int_M^{+\infty} \sqrt{\frac 2{\pi}}\frac{\rho_0\delta_c}M 
\frac 1{\sigma^2} \Big|\frac{d\sigma}{dM}\Big | e^{-\frac{
\delta^2_c}{2\sigma^2}}dM.
$$

We employed the P\&S formalism to find theoretical mass
distributions for all models described in the {\it Introduction
.} Fig.1 shows curves corresponding to 4 models with different
parameters for $\sigma_{16}$ = 0.57.
To calculate $\sigma_R$ we used the approximation for the
transfer function proposed in Pogosyan \& Starobinsky\cite{7}.
Observational data (7 points in total) were extracted from
Bachall \& Cen\cite{8} (the error bars are about 30\%). To find
models satisfying the observations we used the following
criteria: the averaged residues of observational point from P\&S
curve are inside $1\sigma$ for any model and for each point.
Firstly, we found that for each model $\sigma_{16} \in [0.42,
0.58]$ satisfies the observational data. If $\sigma_{16} < 0.42$
the P\&S curve has too small amplitude relatively to
observational data, if $\sigma_{16} > 0.58$ it lies to high. The
result of the calculations for $\sigma_{16} = 0.56, 0.57, 0.58$
is presented at Fig.2. The lines are the upper $1\sigma$ limits.
The allowed region is under the lines. Notice, that the sum of
squared residues is minimal for the model with $n_S = 1.5$,
$\Omega_\nu = 0.1$ ($\sigma_{16} = 0.56$); $n_S = 1.6$,
$\Omega_\nu = 0$ ($\sigma_{16} = 0.57$); $n_S = 1.5$,
$\Omega_\nu = 0$ ($\sigma_{16} = 0.58$). We used $\sigma_{16} 
= 0.57$ as a central 
point by the reasons proposed in White et al\cite{9}. 
Statistically, error bars of data are too large 
to exclude any model even for the bottom line in Fig.2 but we 
argue that there is some tendency to large $n_S$ and small $
\Omega_\nu$.

\begin{figure}[t]
\epsfxsize=9cm
\centerline{\epsfbox{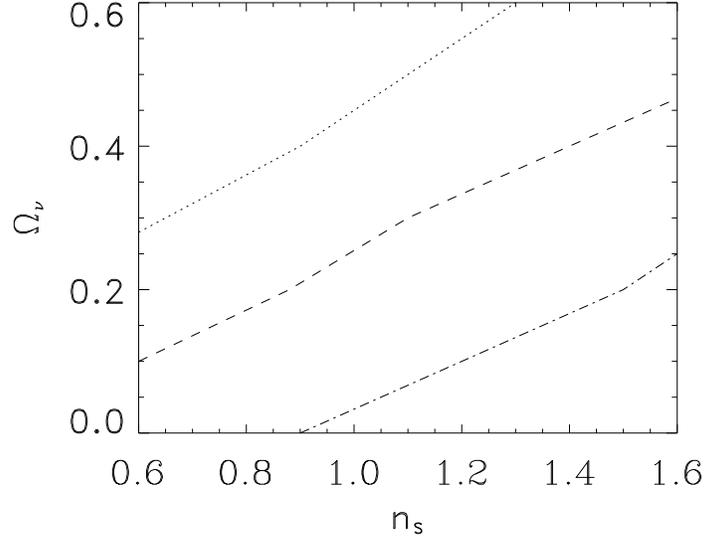}}
\caption{Upper limits for models which are preferable by P\&S 
method for three values of $\sigma_{16}$ parameter. Dotted line 
corresponds to $\sigma_{16} = 0.56$, the dashed line $\sigma_{16
} = 0.57$, the dot-dashed line $\sigma_{16} = 0.58$.}
\end{figure}

\section{$\Delta T/T$}
It is clear that a model should fit the COBE data $\Delta T/T = 
1.06 \times 10^{-5}$, therefore, the simulated amplitude of CMB 
anisotropy should not exceed this value. We simulated $\Delta T
/T\vert_{10^0} \equiv \Delta T/T \vert_{simulated}$ due to
density perturbations and presented the result as slices of 
$\Im \equiv {\rm S}/({\rm S} + {\rm T}) = 
(\Delta T/T \vert_{simulated}) / (\Delta T/T\vert_{COBE})
$ (Fig.3).

We assume that $0.25 \le \Im \le 1$ is a reasonable range that 
does not contradict current theoretical models of inflation 
(Lukash \& Mikheeva\cite{10}). If $\Im > 1$ the CMB anisotropy is 
overproduced, and if $\Im < 0.25$ we have problems with 
constructing the model of inflation. We have obtained that the models 
with $0.75_{\sim}^{<} n_S {}_{\sim}^< 1.25$, $0 \le \Omega_\nu 
\le 0.35$ ($\sigma_{16} = 0.57$) are available only.

\begin{figure}[t]
\epsfxsize=9cm       
\centerline{\epsfbox{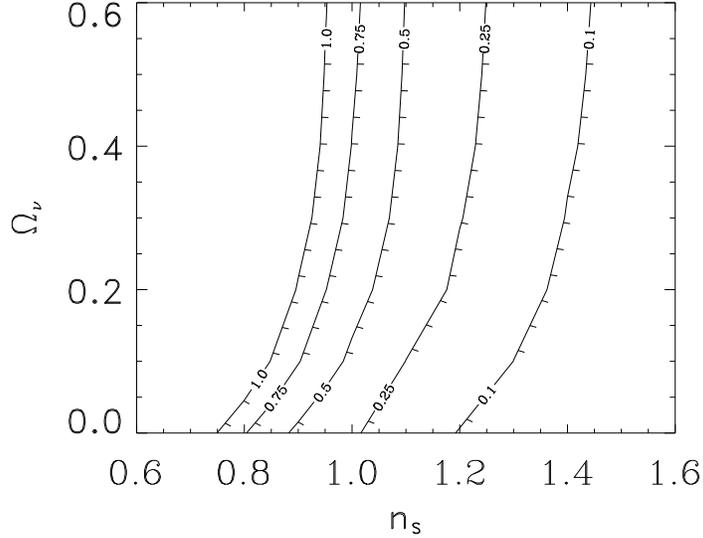}}
\caption{A relative contribution of density perturbations into 
$\Delta T / T \vert_{10^0}$ $\Im$ as a function of model 
parameters $n_S$ and $\Omega_\nu$. $\sigma_{16} = 0.57$. }
\end{figure}

\section {Discussion and conclusions}
Analyzing Fig.2,3 we can conclude (see Fig.4):
\begin{itemize}
\item 
P\&S cluster mass function is strongly sensitive to the $
\sigma_{16}$ value: $\sigma_{16} \in [0.42; 0.58]$;
\item 
the advantaged models by P\&S cluster mass function ($\sigma_{16} =
0.57 \pm 0.01$) are those with the ``blue'' spectra of density
perturbations and any abundance of hot dark matter (see Fig.2);
\item
these ``blue'' models need a significant amount of cosmological 
gravitational waves (T/S$> 1$) to be consistent with the COBE 
data;
\item
Current obsevational data error bars are large, thus $1\sigma$
region covers about a half of the phase space of the models (see
Fig.2);
\item 
for the $\sigma_{16} = 0.57$, $1\sigma$ statistical criterion, 
and $0.25 < \Im < 1$ we obtain the following constrains for the 
model parameters: $0.75^<_\sim n_S {}^<_\sim 1.25$, $\Omega_\nu < 
0.35$ (see Fig.4, the allowable region is shaded);
\item 
if we assume $\Im < 0.25$ then larger $n_S$ (till the boundary 
value 1.6) and $\Omega_\nu$ (till 0.5) are available (see Fig.2).
\end{itemize}

\begin{figure}[t]
\epsfxsize=9cm       
\centerline{\epsfbox{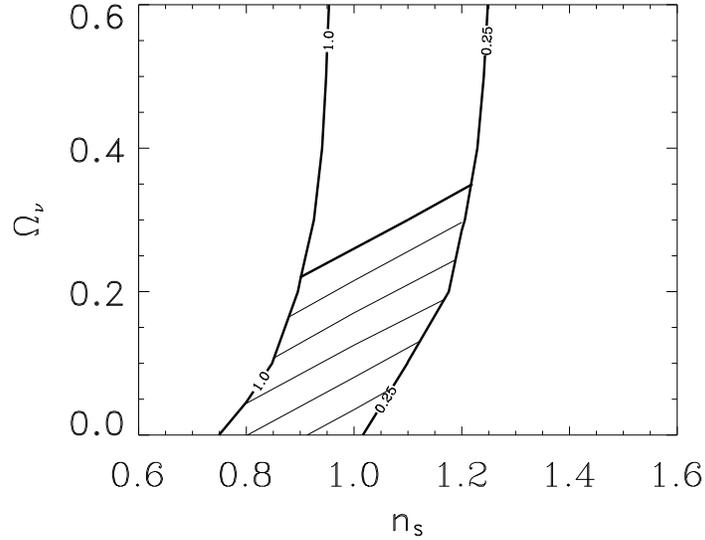}}
\caption{Summarized constraints for power-law spectra of density
perturbations. Labels "1" and "0.25" mark models with T/S = 1 and 
0.25 consequently.}
\end{figure}

{\it Acknowledgements}
The work was partially supported by RFBR (96-02-16689-a), 
COSMION, Russian Federal Programme ``Integration'' (315), and
SNSF 7 IP 050163 (V.N.L. and E.V.M.).


\end{document}